\documentclass[]{ametsocV6.1}
\usepackage{graphicx}
\usepackage{qcircuit}

\title{Quantum Computers for Weather and Climate Prediction: \\ The Good, the Bad and the Noisy.}

\authors{F. Tennie,\aff{a} 
	T. N. Palmer,\aff{a}\correspondingauthor{tim.palmer@physics.ox.ac.uk} 
}

\affiliation{\aff{a}{University of Oxford, Department of Physics, Clarendon Laboratory, Parks Road, \\ Oxford OX13PU, United Kingdom of Great Britain and Nothern Ireland}
}

\abstract{Over the past few years, quantum computers and quantum algorithms have attracted considerable interest and attention from numerous scientific disciplines. In this article, we aim to provide a non-technical, yet informative introduction to key aspects of quantum computing. We discuss whether quantum computers one day might become useful tools for numerical weather and climate prediction. Using a recently developed quantum algorithm for solving non-linear differential equations, we integrate a simple non-linear model. In addition to considering the advantages that quantum computers have to offer, we shall also discuss the challenges one faces when trying to use quantum computers for real-world problems involving "big data", such as weather prediction. } 

\begin{document}

%% Necessary!
\maketitle

%%%%%%%%%%%%%%%%%%%%%%%%%%%%%%%%%%%%%%%%%%%%%%%%%%%%%%%%%%%%%%%%%%%%%
% SIGNIFICANCE STATEMENT/CAPSULE SUMMARY
%%%%%%%%%%%%%%%%%%%%%%%%%%%%%%%%%%%%%%%%%%%%%%%%%%%%%%%%%%%%%%%%%%%%%
%
% If you are including an optional significance statement for a journal article or a required capsule summary for BAMS 
% (see www.ametsoc.org/ams/index.cfm/publications/authors/journal-and-bams-authors/formatting-and-manuscript-components for details), 
% please apply the necessary command as shown below:
%
% Significance Statement (all journals except BAMS)
%
%\statement
%	 Enter significance statement here, no more than 120 words. See \url{www.ametsoc.org/index.cfm/ams/publications/author-information/significance-statements/} for details.
%
%% Capsule (BAMS only)
%%
\capsule
Introduction to quantum computing and review of its potential usefulness for numerical weather and climate prediction.
%       Enter BAMS capsule here, no more than 30 words. See \url{www.ametsoc.org/index.cfm/ams/publications/author-information/formatting-and-manuscript-components/#capsule} for details.
% 
%% * * If using twocol mode, you will need to use the commands "twocolsig" and "twocolcapsule" in place of "sig" and "capsule"
%%      to ensure that the text box correctly spans across both columns.

%%%%%%%%%%%%%%%%%%%%%%%%%%%%%%%%%%%%%%%%%%%%%%%%%%%%%%%%%%%%%%%%%%%%%
% MAIN BODY OF PAPER
%%%%%%%%%%%%%%%%%%%%%%%%%%%%%%%%%%%%%%%%%%%%%%%%%%%%%%%%%%%%%%%%%%%%%
%
\section{Introduction}
Hardly a day goes by without the announcement of some new breakthrough in quantum technology. Quantum computers can already perform calculations that classical computers would be unable to perform in reasonable lengths of time - quantum supremacy as it is known. Although current quantum computers have limited power, there are plans for next generation quantum computers with millions of so-called qubits - the elements of a quantum computer. 

This paper was motivated by a comment made by one of our colleagues: ``As soon as we have a quantum-fortran compiler, we'll be able to start using these computers in meteorology.'' Unfortunately it is not as simple as this - the way quantum computers work is profoundly different to the way classical computers work. Indeed it is so different that it is very difficult for someone who is not well versed in quantum mechanics to understand what exactly these differences are. 

In this paper, we try to answer the question of whether quantum computers will ever take over from classical computers in trying to make weather forecasts, or seasonal predictions, or indeed climate-change projections, using the underlying laws of physics. We do this assuming a typical meteorological audience with a good knowledge of fluid mechanics and classical computers but little knowledge of quantum computing theory. 

In attempting to answer this question, we have been motivated by the title of the movie ``The Good, the Bad and the Ugly", but replacing it with the The Good, the Bad and the Noisy. In quantum computing there are certainly good points and bad points. The question is which will win out overall. In this review, we have taken the opportunity to discuss some advances of our own, using quantum algorithms to solve non-linear differential equations exponentially faster than classical algorithms.

We do not aim to provide a technical review article on the field of quantum computing. For a comprehensive introduction the interested reader should consult the excellent book by \citet{Nielsen2010}. A selection of reviews on various quantum computing paradigms and recent developments may be found in \citet{Nimbe2021}, \citet{Bharti2022}, \citet{Montanaro2016}, and \citet{Cerezo2021}.

\section{The Good}\label{Sec:GOOD}

\subsection{Key elements of Quantum Computation}

Everyone knows that the elements of a classical computer are its bits. Each bit has a value of 0 or 1. The elements of a quantum computer are qubits. It is easy to describe what a qubit is since it has a precise mathematical definition. However, it is difficult to describe what a qubit means physically since (it is safe to say), the physical meaning of quantum states is still a matter of debate and discussion. Some pedagogical discussions of quantum computing try to find simple analogies for qubits, in an effort to make them understandable. However, the truth of the matter is that they are not, currently at least, understandable in terms of intuitive analogues. In the following, it is probably best to follow Richard Feynman's dictum 

\begin{quote}Do not keep saying to yourself, if you can possibly avoid it, ``But how can it be like that?'' because you will get `down the drain', into a blind alley from which nobody has escaped. Nobody knows how it can be like that.
\end{quote}

So instead, we will describe a qubit mathematically. It is an element of a complex Hilbert Space: a vector space equipped with an inner product where the vectors have complex amplitudes. Just as in physical space, a vector can be described by its components relative to an orthogonal basis, i.e. 
\begin{equation}
\mathbf{v}= a\; \mathbf{\hat {i}} +b \; \mathbf{\hat j} + c \; \mathbf{\hat k}
\end{equation}
where $a$, $b$ and $c$ are real numbers, so a vector in a 2-dimensional complex Hilbert Space can be written 
\begin{equation}
\label{1q}
|\psi\rangle = \alpha \; |0\rangle + \beta \; |1\rangle
\end{equation}
where $\alpha$ and $\beta$ are complex numbers. In this representation, the orthogonal basis vectors $|0\rangle$ and $|1\rangle$ are completely arbitrary, just as $\bf \hat i$, $\bf \hat j$ and $ \bf \hat k$ are arbitrary. With 2 complex numbers, it would seem that this vector has 4 degrees of freedom. However, in quantum mechanics, the vectors are normalised, and there is an undefined global phase factor, meaning that in fact there are actually only 2 degrees of freedom. In the appendix, we provide further details on qubits, how they compare to classical bits and on the Dirac notation used in Eq.~(\ref{1q}).

The very simplest quantum computer (too simple to be useful) would take as its input state the vector $|\psi\rangle$ in Eq.~(\ref{1q}) with some initial values $\alpha(t_i)$ and $\beta(t_i)$. There would then be a quantum circuit where certain rotations of the vector are performed in hardware. At the end of these rotations, the vector is still in the state (\ref{1q}) but with final values $\alpha(t_f)$ and $\beta(t_f)$. These rotations are effected by the governing equation of quantum mechanics, the linear Schr\"{o}dinger equation. In terms of the complex Hilbert Space, the operations that map initial vectors to final vectors can be described by unitary matrices. 

There is one final, yet utterly critical part of our simple quantum computer. One has to ``measure'' the qubit. Again, whilst there is an experimentally well defined procedure for performing such measurements, no-one is quite sure what is going on at a theoretical level when a measurement takes place. For example, 2020 Nobel Laureate Roger Penrose has proposed that the process of measurement occurs when the effects of self gravitation of quantum systems are no longer negligible. 

Again, we ignore such deep matters, and simply describe the recipe which all students of quantum mechanics know. To perform the measurement, you first choose a vector basis relative to which the measurement takes place. If $|0 \rangle$ and $|1\rangle$ describe this measurement basis, then, after measurement, the qubit will be found to be in the $|0 \rangle$ state with probability $|\alpha(t_f)|^2$ and in the $|1\rangle$ state with probability $|\beta(t_f)|^2$. This appearance of probability led Einstein to complain that God does not play dice. However, most physicists today believe that quantum uncertainty is fundamentally encapsulated in the laws of quantum physics, making it different in character to chaotic uncertainty (whose origin is our inability to know about flaps of butterflies wings). One of the authors believes that scientific orthodoxy will ultimately be overturned on this matter (\cite{Palmer2022}). However, this is not the topic of this paper. 

In practice, it can be arranged so that a bell rings, or a light flashes, when the state evolves to $|0 \rangle$. We can estimate the modulus of the complex amplitudes by carrying out multiple measurements of the state and counting the number of times the state evolved to $|0\rangle$. 

\begin{figure}
	\centering
	\includegraphics[width=0.7\linewidth]{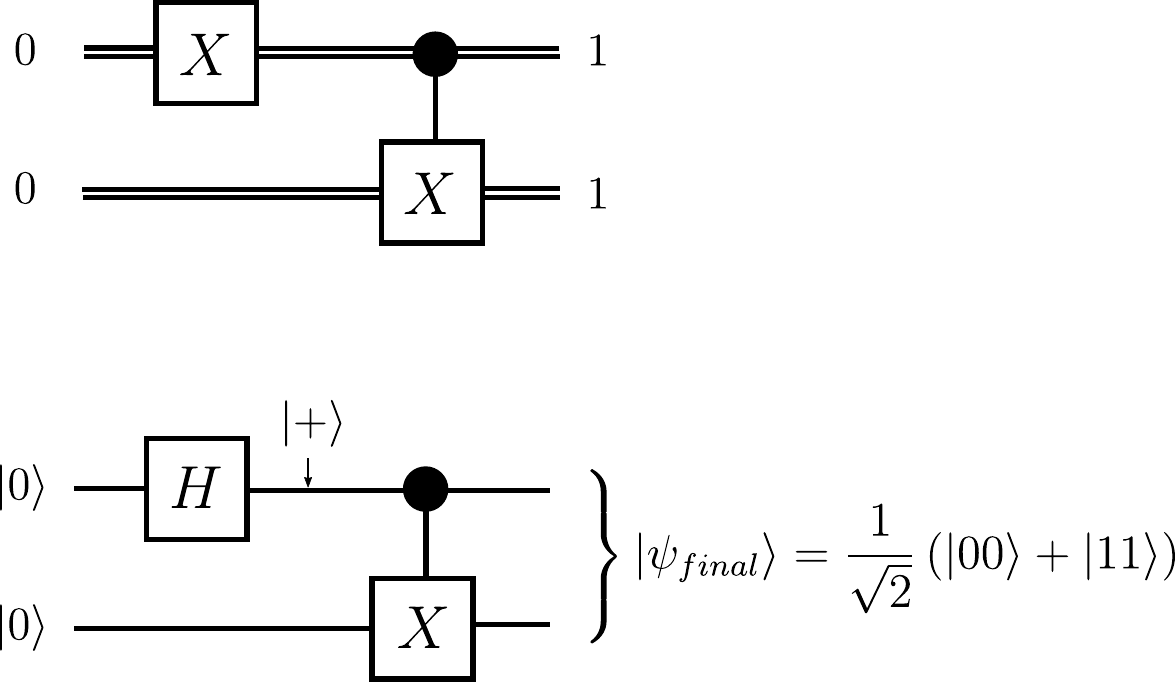}
	\caption{Top: Classical circuit processing two bits initialised in the state '0'. The X-gate flips the upper bit and the controlled-X-gate subsequently flips the lower bit. Throughout, each bit is in a definite state of either '0' or '1'. Bottom: Simple quantum circuit. The Hadamard-(H)-gate transforms the upper qubit into the state $|+\rangle = \frac{1}{\sqrt{2}}(|0\rangle + |1\rangle)$. The controlled-X-gate then entangles both qubits. The final state corresponds to $\gamma_1=\gamma_4=\frac{1}{\sqrt{2}}$ and $\gamma_2=\gamma_3=0$ in Eq.~(\ref{2q}). Note that the information 'stored' in the amplitudes is non-local and cannot be attributed to an individual qubit unlike in the classical case.}
	\label{fig:examplecircuit}
\end{figure}

The power of a quantum computer comes from being able to `entangle' qubits together. We form multi-qubit states by what are called tensor products of Hilbert vectors. For example, the Hilbert State of a two qubit system can be written
\begin{equation}
\label{2q}
|\psi\rangle = \gamma_1 \; |00 \rangle + \gamma_2 \; |01\rangle + \gamma_3 \; |10\rangle + \gamma_4 \; |11\rangle .
\end{equation}
Here there are now four basis states, and on measurement the system evolves to one of these basis states with a probability given by the squared modulus of the associated complex amplitude. Fig.~\ref{fig:examplecircuit} illustrates and contrasts a 2-bit classical and a 2-qubit quantum circuit. 

Now we can already start to glimpse the reason why quantum computers are powerful. 2-qubit states are associated with 4 complex amplitudes, i.e. 8 real numbers (less 2 for normalisation and global phase). It's easy to generalise (\ref{2q}). A $n$ qubit state can be written
\begin{equation}
\label{mq}
|\psi\rangle = \alpha_1 \; |\underbrace{00\ldots 00}_n\rangle + \alpha_2 \; |\underbrace{00 \ldots 01}_n\rangle + \alpha_3 \; |\underbrace{00 \ldots 10}_n\rangle +\ldots + \alpha_{2^n} \; |\underbrace{11 \ldots 1}_n\rangle .
\end{equation}
In this way, $n$-qubit states are associated with $2^n$ complex numbers, i.e. $2^{n+1}$ real numbers (less 2 for normalisation and global phase). This exponential increase in the dimension of the Hilbert Space with qubit number $n$ has some remarkable consequences. 

Consider a modern numerical weather prediction (NWP) model, with say 1 billion variables. These variables represent the components of wind velocity, temperature, humidity, and other variables, on a grid that covers the Earth and extends up into the stratosphere. How many qubits are needed to encode the state of such an NWP model into the complex amplitudes of a multi-qubit Hilbert state? A simple estimate is to take the log to base 2 of 1 billion. The answer is 30. By modern quantum computer standards, a 30-qubit quantum computer is not an especially large computer. 

Hence, it appears possible to initialise a 30-qubit quantum computer with the initial state of a billion-variable NWP model. So far, so good! The next question is whether it is possible to encode the dynamical equations of a NWP model (e.g.~the Navier-Stokes equation) into the unitary transformations that a quantum state evolves through before measurement. It may seem there is a fundamental obstacle here - that the Schr\"{o}dinger equation is linear, whilst the Navier-Stokes (and other equations in NWP) are non-linear. However, in fact that obstacle can be overcome. This is an area of research in which the authors, in collaboration with a group at MIT, are actively engaged. In the subsection below we give a brief summary of this research. As with much in quantum computing, it is impossible to provide a schematic outline without having to give some techincal details. The reader not interested in such details can skip to the end of this section.

In addition to the quantum computing paradigm described above which consists of a controlled unitary evolution of quantum states followed by measurements of the qubits, a range of other approaches have been suggested. A type of quantum machine referred to as an annealer, such as the much reported D-Wave machine, seeks to encode computational tasks as minimisation problems of a target function (\citet{McGeoch2018}). By carefully tayloring the interaction between the internal qubits, the solution of the minimisation is given by the lowest energy state of the machine. It is yet unclear whether this approach offers a quantum advantage and whether it is computationally universal. 

Hybrid quantum computing represents yet another quantum computing paradigm, where the quantum hardware is used to evaluate a parametrised target function. The overall task is to find the parameter values which minimise the target function. After each evaluation of the target function on the quantum chip, a classical computer determines a new set of parameters for which the target function will be evaluated until a local optimum has been determined. As with the D-Wave machine, it has yet to be shown that this scheme can offer a scalable quantum advantage. 

\subsection{Non-linear Differential Equation Quantum Solver}

Our way to solve non-linear differential equations is to encode multiple copies of a quantum state and let these copies interact with each other - via what in quantum theory is known as an interaction Hamiltonian (cf.~\cite{QuantSolver2020}). We then study the evolution of one of the copies. It turns out that the equation of motion for a single copy is itself non-linear; the non-linearity originates as the leading order term in the expression for the interaction between copies. It can be understood and described as an effective mean-field that arises from averaging the degrees of freedom of all but one copy. In this framework, however, one finds that higher order terms give rise to mean-field errors. These errors tend to accumulate during the course of a computation, which will be demonstrated in an example further below.  In a recent paper, Lloyd et al.~(2020) utilised mean-field non-linearities to create a quantum non-linear solver that offers an exponential resource gain over classical solvers. We give a brief description of this quantum algorithm, based on the ideas above, in the Appendix.

\begin{figure}[h]
	%	\begin{subfigure}[b]{0.48\textwidth}
	%		\includegraphics[width=\textwidth]{1d_CubicModel_DifferentInitialValues}
	%		%\caption{Picture 1}
	%		\label{fig:1}
	%	\end{subfigure}
	%	%
	%	\begin{subfigure}[b]{0.49\textwidth}
	%		\includegraphics[width=\textwidth]{1d_CubicModel_DifferentCouplingValues}
	%		%\caption{Picture 2}
	%		\label{fig:2}
	%	\end{subfigure}
	\includegraphics[width=\textwidth]{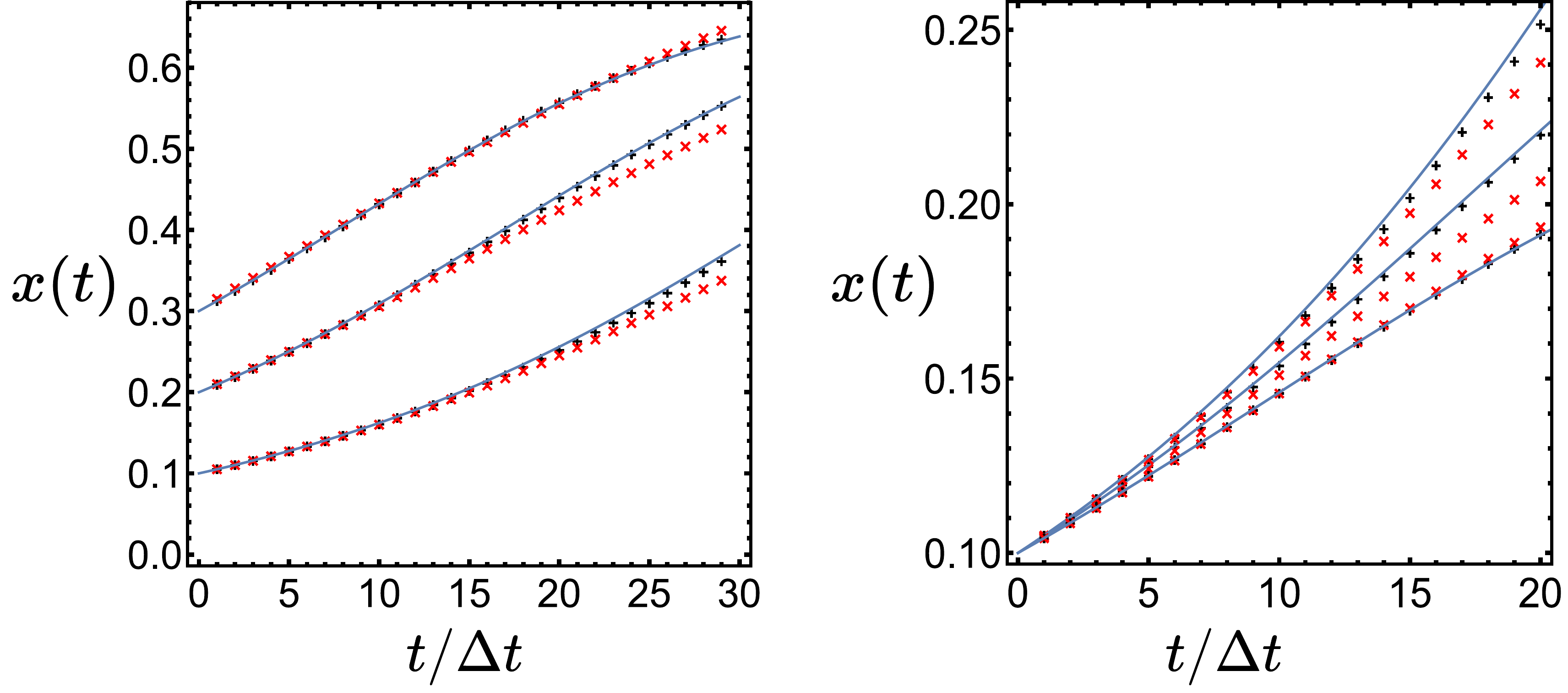}
	\caption{Results of integration of $\dot{x} = x - \alpha x^3$. Red diagonal crosses mark the output of the quantum solver; by comparison the forward Euler method executed on a classical computer results in numerical values marked by black horizontal/vertical crosses. The blue solid line represents the analytic solution. Left: integration of the model for different initial conditions $x_0 = 0.1, \,0.2,\, 0.3$ for a relative coupling strength $\alpha=2$. Right: integration of the model for different coupling strengths $\alpha = 2,8,16$ (top to bottom branch) and initial value $x_0 = 0.1$. In both cases the step size was chosen as $\Delta t = 0.05$ and the quantum solver was initialised with $N=15$ identical copies of the state.}\label{model1}
\end{figure}

For the rest of this section we are going to apply the Lloyd et al algorithm to a simple non-linear differential equation. Consider the flow of a single dynamical variable subject to $\dot{x}(t)=x - \alpha x^3$. For any finite initial value $x_0$, the system will converge to a fixed value in a sigmoide-shape trajectory in the $t$-$x$ plane. We have emulated the execution of the quantum solver on a classical computer because the quantum resources currently available are not yet sufficient to run the solver (see discussion below). The results are shown in Fig.~\ref{model1}.

In the left figure, the output of the quantum solver is shown for various different initial values. It is benchmarked by a classical forward Euler integration and the analytic solution. We observe that the output of the quantum solver is in good agreement with the classical output up to about $15$ integration steps. This is no mere coincidence. In carrying out the integration, $N=15$ identical system copies have been used to run the quantum solver. As mentioned above, the inclusion of non-linear terms leads to additional mean-field errors which build up during each integration step. As those errors usually are surpressed by a factor of $1/N$, the errors limit the range of the integration to $\mathcal{O}(N)$ steps. Although this might appear to be a draw-back of the quantum solver, the linear scaling actually represents a major improvement over former quantum algorithms used to integrate non-linear systems, which required $\mathcal{O}(Exp[N])$ copies, cf. \citet{Leyton2008}. In the second figure, we can see that the quantum solver is capable of handling various different relative sizes of the non-linear terms. Generally, however, we expect that the number of required system copies increases with the relative strength of the non-linearities of the system. Unstable systems with positive Lyapunov expoents might require an exponentially increasing number of copies.

\begin{figure}[h]
	\hspace{2.7cm}
	%	\begin{subfigure}[b]{0.46\textwidth}
	%		\includegraphics[width=\textwidth]{Inward_2d_spiral}
	%		%\caption{Picture 1}
	%		\label{fig:1}
	%	\end{subfigure}
	%
	%	\begin{subfigure}[b]{0.47\textwidth}
	%		\includegraphics[width=\textwidth]{Inward_2d_spiral_zoom_variousDelta}
	%		%\caption{Picture 2}
	%		\label{fig:2}
	%	\end{subfigure}
	\includegraphics[width=0.6\textwidth]{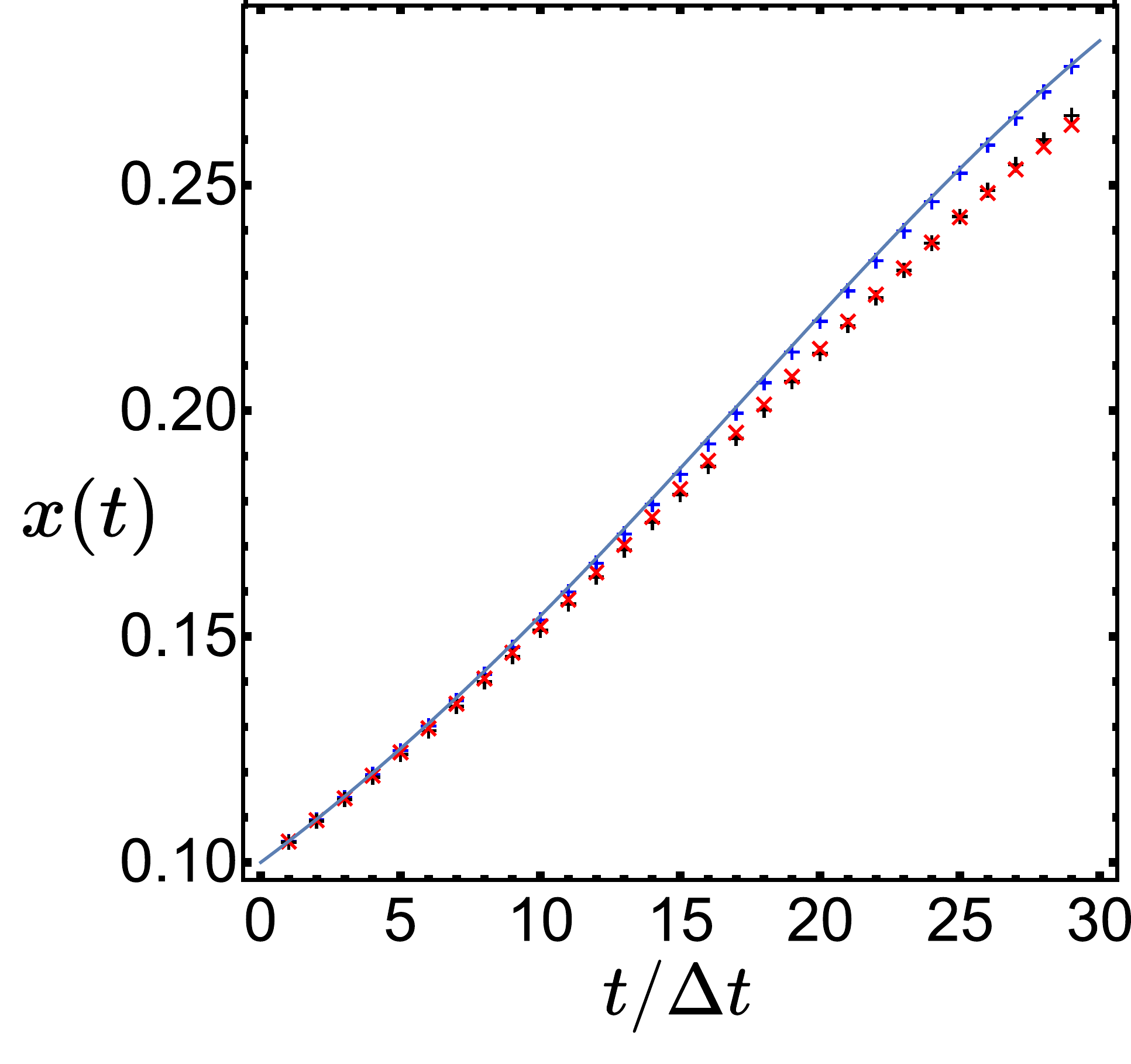}
	\caption{Integration of $\dot{x}=x-8 x^3$. Comparison of a classical forward Euler solution (blue horizontal/vertical crosses) with an ensemble-averaged solution (black horizontal/vertical crosses) and the quantum solver output (red diagonal crosses). The solid blue line represents a Runga-Kutta method based solution. The step size was chosen as $\Delta t = 0.05$ and the quantum solver was initialised with $N=10$ identical copies.}
	\label{model2}
\end{figure}

Classical NWP utilises ensemble methods (cf.~\citet{Palmer2022}). It is therefore instructive to compare the results of integrating our one dimensional toy model using the quantum solver to a classical ensemble-average solution. As shown in Fig.~\ref{model2}, we find that the quantum solver output is closer to the ensemble average than to a single classical forward Euler solution. If generally true, this would be a remarkable result. Further research will be required to establish whether this observation holds true for arbitrary dynamical systems. 

An obvious next step would be to integrate the famous Lorenz (1963) equations with our quantum algorithm. However, there is a difficulty on classical emulators of quantum computers which we are currently working to overcome when the time derivatives are expressed as polynomials of even degree. At this point, we have determined that the problem can be solved by supplying the single copy state vector with additional ``dummy" variables. The number of these dummy variables will scale linearly with the number of integration steps $T$. Although this does not present a problem for a quantum computer per se, it does render the emulation of the quantum solver on a classical computer unfeasible. 

One might therefore ask at which point quantum hardware will become able to run the quantum solver algorithm. Let us assume that we want to integrate the Lorenz system for $T=100$ integration steps. This corresponds to inverting a matrix of size
\begin{equation}
N \approx (\# \text{dynamical variables} \cdot T )^{T} \approx 10^{200}
\end{equation} 

The state vector of size $N$ can be encoded in $\log_2[10^{200}]\approx 655$ qubits. In addition, there is a number of ancilla qubits linked to the precision of the matrix inversion which can be estimated to be about $100$. Thus, we expect a required circuit width of a few hundred error corrected qubits to be sufficient for addressing the Lorenz system. Currently available quantum hardware such as IBMQ offers about 100 noisy qubits with plans to achieve about 1000 in the coming years. (For a discussion on the impact of noise, cf.~Sec.~\ref{sec:noisy} .) 

So far, it thus sounds plausible to expect implementation of the quantum solver in a few years time. Another important quantity, however, is the circuit depth, i.e. the number of required elementary gates. Here, it is more difficult to predict the required resources. \citet{Scherer2017} predicted a circuit depth of $10^{29}$ gates to invert a matrix of size $N\approx 3*10^9$ in the context of scattering cross section computations. If this is correct, this poses a twofold problem. First, assuming that execution times of the individual gates are of the order of $10^{-9}s$, the runtime of the algorithm would be about $10^{20}s$. Second, in order to keep control of the errors in gate execution, a single gate would need to have a fideliy of $\mathcal{F}\approx 1 - 10^{-29}$. This is not realistic. Note that this gate count is closely related to properties of the matrix to be inverted. For instance, a recent publication by \citet {Perelshtein2022} reported on the inversion of a matrix of size $N = 2^{17}$ on currently available quantum backends from IBM. 

In summary, the circuit depth required for the integration of the Lorenz system crucially depends on parameters such as the connectivity of future quantum hardware. The authors of this paper believe that the benefit of their work lies in informing the design of future quantum hardware such that computational overhead will be kept to a minimum.

\section{The Bad}

Readers may be familiar with the parlor game `20 Questions'. One person thinks of some kind of object and the rest of the group have to guess what the object is by asking questions. However, the question must be such that it can only be answered with a `yes' or a `no'. Hence, if it has been ascertained that the object is a piece of fruit, you aren't allowed to ask `What type of fruit is it?'. Instead you must ask `Is it an orange?'. 

Something like these two types of question describe well the difference between classical physics and quantum physics. On a classical computer, if you want to know the value of some particular variable, such as temperature in gridbox N at timestep M, you simply ask the computer to write out that value. On a quantum computer, things are much more nuanced. You might start by asking the quantum computer if the temperature (at gridbox N and timestep M) is greater or less than 290K. To do this we have to add to the quantum hardware a unitary operation such that if the temperature is greater than 290K then the state evolves to the `$|0\rangle$', so that, if this state is measured, the alarm bell, or green light, or whatever, will light with complete certainty. 

Here, the fact that the state space of $n$ qubits is exponential in $n$ comes back to bite us. The converse of this remarkable exponential fact of quantum life is that the number of qubits that we can measure, measurements that are needed if we are to discover the results of the unitary transformations between the initial and evolved quantum states, is the logarithm of the dimension of that state space. 

If we have a 30-qubit system, then essentially we can ask 30 questions about the weather around the world. Is it raining in Rio, tomorrow? Is it stormy in San Francicso next week? Is the monthly outlook for Copenhagen cold?

If we had a proprietorial quantum computer and all we cared about was the weather in our back yard, then perhaps we would be satisfied with such limited information. Perhaps for climate change, all we care about is whether the global-mean temperature will rise above 2C. However, of course, for any realistic operational weather or climate center, this is far from sufficient. We need to know the value of the many variables in an NWP model - maybe not all billion of them, but certainly millions of values for the many variables at many parts of the world at many times. 

Since the output of a quantum computation will be limited to as many bits as it processes qubits, this is an (insurmountable) bottleneck. Therefore, quantum computers are unlikely to entirely replace classical computers for NWP. They may however play a role performing specialised tasks within an NWP or climate model. This is consistent with the trend towards more heterogeneous computing architectures (e.g.~CPU vs GPU). Quantum computers might offer an advantage in answering questions that require a resolution exceeding classical resources. However, we are a very long way from having both quantum algorithms and quantum hardware that can accomplish such specialised tasks. Indeed, our main conclusion is that quantum computers are not well suited to ``big-data" problems, where outputs explore the high-dimensionality of the big data.

\section{The Noisy}\label{sec:noisy}

In the first section we have seen that we can encode the state of an NWP model onto a  quantum computer and that we can perhaps describe the Navier-Stokes and other non-linear equations of NWP with the Schr\"{o}dinger equation, through an interaction Hamiltonian. In this section, we jump to what seems like an `ugly' aspect of quantum computing, but argue it may not be as ugly as we might imagine at first sight. 

One of the problems with quantum entanglement is that quantum systems love to get entangled with their environment. However, the environment is typically a messy, noisy system. Hence, when qubits entangle with the environment, they themselves become noisy. Quantum computer manufacturers try to minimise such environmental `decoherence' by enclosing the quantum computer in a supercold superconducting environment. However, this only works to a degree. 

As a result, we have to distinguish between hypothetical noise-free `logical qubits' on the one hand, and practical noisy `physical qubits' on the other. It is possible to represent these hypothetical logical qubits if you have enough copies of the noisy physical qubits, by applying error detection and correction algorithms to the copies. 

So how many copies do we need? Well the answer depends on what level of accuracy and precision we require in the first place. Furthermore, one might be able to take advantage of the fact that in NWP on classical computers, there are considerable advantages to introducing noise into computational representations of the non-linear Navier Stokes equation (cf. \citet{Palmer2019}). Is there a quantum analogue? Due to interaction with the environment, the evolution of noisy qubits during a computation must be described by a more generalised formalism than the Schr\"{o}dinger equation. While so-called density matrices are useful in understanding the outcome of several averaged runs of a noisy quantum computer, not-averaged noise encountered in a single run can be captured by the stochastic Schr\"{o}dinger equation (\citet{Bouten2004}). Here, the evolution of the quantum state in a single run of the quantum computer is subject to stochastic Wiener processes. Although the precise form of the stochastic terms critically depends on the design of the quantum hardware, it is plausible that stochastic terms will be injected into the step-wise integration of non-linear systems when executed by the quantum solver. Further research will be required to establish whether potential future quantum NWP could actually benefit from this feature.

A new insight which resulted from writing this paper is that the integration of stochastic differential equations can in fact be accomplished using an extended form of the quantum solver. Here, the noise terms present at each integration step are inserted during the initial set-up of the quantum computer. This means that the noise does not have to be injected at  each discretised intergration step, but will rather be a part of the initial quantum state.

Early studies (\citet{Epstein1969}, \citet{Fleming1971a}, \citet{Fleming1971b}) of the state-dependent predictability of weather focussed on solving the classical Liouville equation (so-called stochastic-dynamic prediction). However, in practice this proved impractical due to the large dimensionality of the state space. It is currently unclear whether noisy quantum computers would offer any advantage in solving the classical Liouville equation for such weather forecast problems - at present there is no evidence that it would.

The introduction of noise can help to reduce the dimensionality of a dynamical system. For instance, a suitable variable transformation of the deterministic three-dimensional Lorenz system yields a two-dimensional stochastic system (\citet{Palmer2001}).
This offers a potential route to gauge the impact of noise in a quantum computer using near-term quantum hardware: by integrating the modified Lorenz system of reduced-dimensionality on the quantum solver, the imperfections of the hardware might allow to explore chaotic behaviour of the -effectively- stochastic system and to reproduce the Lorenz attractor.

%DOES IT MATTER WHETHER NOISE IN NWP IS PSEUDORANDOM OR TRUELY RANDOM? THE METHOD BRIEFLY MENTIONED IN THE LAST TWO SENTENCES USES CLASSICAL NOISE GENERATED BY A CLASSICAL COMPUTER. LOTS OF PEOPLE STRESS THAT QUANTUM NOISE IS VERY DIFFERENT AND REPRESENTS TRUELY RANDOM NOISE. (REAL VS PSEUDORANDOMNESS). 

\section{Conclusions}

In writing this paper, the authors hope to shed light on the potential applicability of (future) quantum computing technologies in weather and climate predictions. By making use of an exponentially large state space, quantum computers offer a range of advantages when it comes to processing large sets of data. Recent developments have paved the way to integrating non-linear systems on quantum computers (\citet{QuantSolver2020}). We have emulated the behaviour of a non-linear quantum solver and presented encouraging results for a number of toy models. 

Unfortunately, quantum computing does not only offer benefits. In the authors' opinion, the two main challenges for quantum computers to become useful tools for researchers of disciplines involving numerical computations are the limited amount of information that can be read out after a computation and noise. While a plausibility argument seems to indicate that a certain level of noise could even be beneficial in the integration of non-linear dynamics, the limitations on the read-out capacity present a seemingly insurmountable problem. However, in some quantum computing paradigms such as, e.g., hybrid quantum computing, reading out the full quantum state might not be required. Unfortunately, it is yet unknown whether in these frameworks there will be a scalable quantum advantage.

%In hybrid quantum computing, a parameter dependent loss function is evaluated by a quantum computer while a classical computer optimises the parameters. At the time of writing this paper, no hybrid algorithm is known to the authors that provides a verifiable quantum speed up when integrating non-linear systems. If no further theoretical insights should be established, the potential of hybrid schemes could thus only be gauged by running large simulations on future quantum hardware.

With the development of more powerful quantum processors in the upcoming years, it is likely that quantum computers will become useful tools for a range of specialised applications. They, however, are unlikely to replace classical computers for NWP or climate prediction in the foreseeable future. 

\section{Data Availability Statement}

Due to confidentiality agreements, supporting data can only be made available to bona fide researchers subject to a non-disclosure agreement. Details of the data and how to request access are available from felix.tennie@physics.ox.ac.uk.

\appendix

\section*{Bits vs. Qubits}

Classical bits are variables assuming one of two definite values such as $0,1$ or $\uparrow,\downarrow$. Each bit may be regarded as a representation of the cyclic group $Z_2$ and can be physically implemented in any system possessing two (macroscopically) distinguishable states such as, for example, the local (macroscopic) magnetisation of a tape. If we label the two states of a bit by computational basis vectors of a 2-dimensional vector space, the system can either be in state
\begin{equation}
	\begin{pmatrix}
		1 \\
		0
	\end{pmatrix}\sim\, \uparrow \text{     or     }
	\begin{pmatrix}
		0\\1
	\end{pmatrix}\sim\,\downarrow.
\end{equation}
Multiple bits form bit strings. At any point during a computation on a deterministic classical computer those bit strings are in a definite state, e.g. $\overline{b}=\overline{0101100}$.

\begin{figure}
	\includegraphics[width=\linewidth]{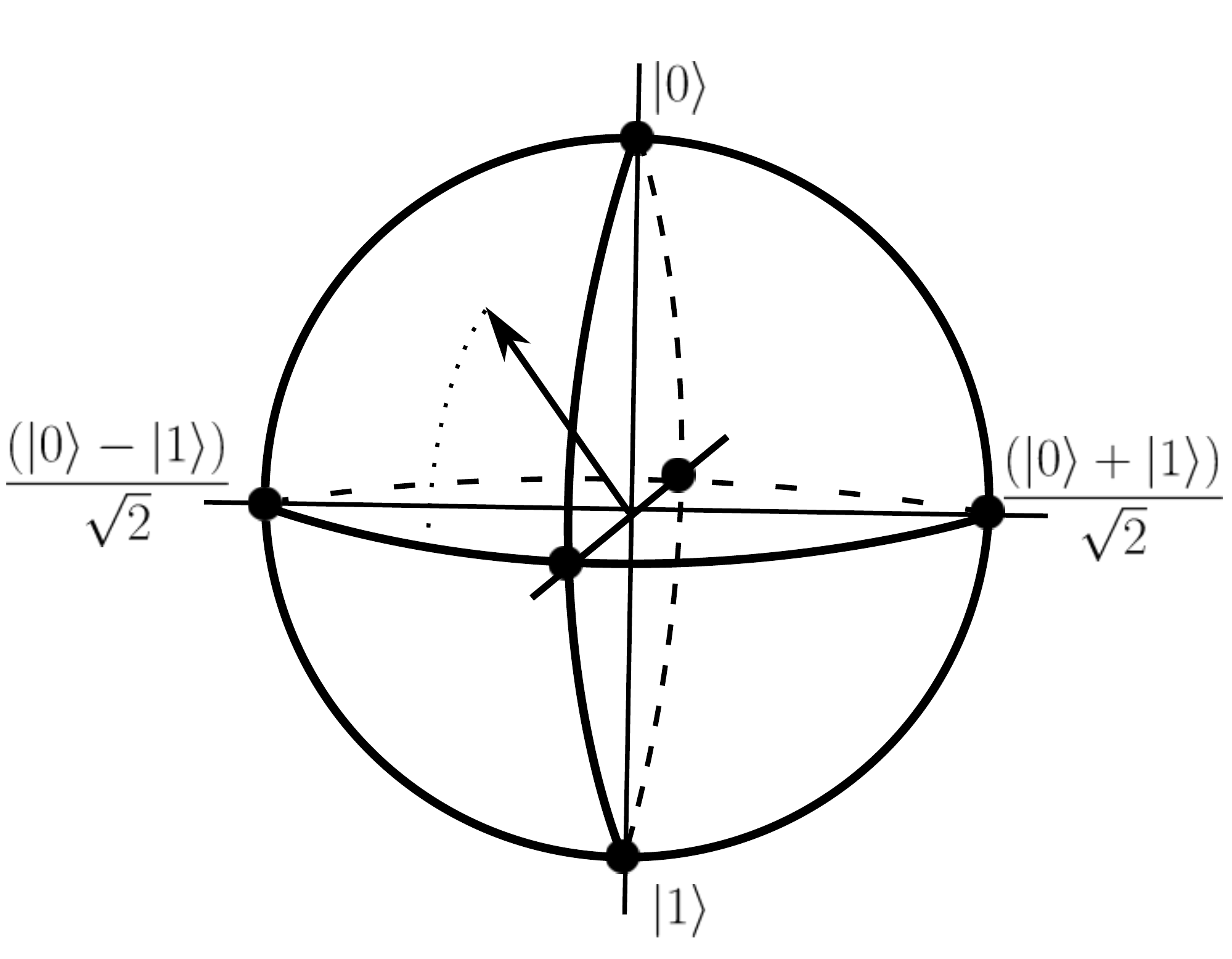}
	\caption{Bloch sphere  representing the set of all qubit states. The states $|0\rangle$, $|1\rangle$ lie at cartesian coordinates $(0,0,1)$ and $(0,0,-1)$ respectively. The states on the x-axis have no classical analogue.}
	\label{fig:BlochSphere}
\end{figure}

The fundamental unit of quantum information is the \textit{qubit}. It describes the state of a microscopic 2-level system such as a spin-$\frac{1}{2}$-particle and can be represented by a complex vector
\begin{equation}
	\psi = \alpha \begin{pmatrix}
		1 \\
		0
	\end{pmatrix}+ \beta
	\begin{pmatrix}
		0\\1
	\end{pmatrix},
\end{equation}
where $\alpha,\beta \in \mathbb{C}$ and $|\alpha|^2 + |\beta|^2=1$. Commonly, this is written in Dirac's bra-ket notation as\footnote{Henceforth, we shall use $0,1$ and $\uparrow,\downarrow$ synonymously.}
\begin{equation}
	|\psi \rangle = \alpha |0\rangle+ \beta |1\rangle.
\end{equation}
The set of qubits can be visualised as a sphere where $\alpha = cos(\theta/2)$ and $\beta = sin(\theta/2)e^{(i\phi)}$ for spherical coordinates $\theta$ and $\phi$ as illustrated in Fig.\ \ref{fig:BlochSphere}. The complex phase of $\alpha$ may be chosen as zero since it has no physical meaning for a single qubit.

For non-vanishing $\alpha,\beta$ the qubit is said to be a coherent superposition of the two distinguishable states associated with a classical bit. Note that this has no equivalent in the classical macroscopic viewpoint. When the qubit gets measured, i.e.\ evaluated, the amplitudes $\alpha$ and $\beta$ determine the propbability of either outcome $"0"$ or $"1"$ by virtue of the Born-rule:
\begin{equation}
	P(\uparrow) = |\alpha|^2 \text{     and      }P(\downarrow) = |\beta|^2.
\end{equation}

Commonly, the probablistic nature of the read-out process causes potential for confusion. However, note that a qubit is not simply a classical non-deterministic bit. Although such a non-deterministic bit has a certain probaility to assume either value $0$ or $1$, it cannot be used to describe a system such as an electron-spin. The space of all states of a non-deterministic bit may be visualised as the interval $[0,1]$ which is clearly not isomorphic to the hull of a sphere representing all states of a qubit. For example, a non-deterministic bit with $P(0)=P(1)=0.5$ cannot capture the notion of the two \textit{physically} different states
\begin{equation}
	|+\rangle\equiv \frac{|0\rangle + |1\rangle}{\sqrt{2}} \text{        and        }|-\rangle\equiv \frac{|0\rangle - |1\rangle}{\sqrt{2}}.
\end{equation}
Although both states $|+\rangle$ and $|-\rangle$ give rise to the same probabilities for measuring $0$ or $1$ (i.e. $p=0.5$), they behave very differently when manipulated further as will be seen in subsequent sections.

The general quantum state of $n$ qubits can be described as an element in a complex vector space of dimension $2^n$. For instance, the general two-qubit state may be written as:
\begin{equation}\label{eq:GeneralTwoQubitExpansion}
	|\Psi \rangle = c_1 |0\rangle \otimes |0\rangle + c_2 |1\rangle \otimes |0\rangle + c_3 |0\rangle \otimes |1\rangle + c_4 |1\rangle \otimes |1\rangle,
\end{equation}
where $\sum_{i}|c_i|^2=1$. Note the outer products of single-qubit states forming the basis of the two-qubit state space.\footnote{In the literature, the short form $|i j\rangle$ is often used for $|i\rangle \otimes |j\rangle$, etc.}
Again, the probabilities of finding the two qubits in a particular state are determined by the Born rule. For instance, the joint probability of measuring the first qubit in the state $0$ and the second in the state $1$ is given by $|c_3|^2$.

A crucial difference to classical bits can be observed when quering the state of a single qubit in a multiple qubit state. In a classical bit string, each bit has a well-defined value, e.g. the second bit of $\overline{b}=\overline{1011}$ is in the state $0$. Consider now the two-qubit state
\begin{equation}
	|\Phi\rangle = \frac{1}{\sqrt{2}}(|0\rangle \otimes |0\rangle +|1\rangle \otimes |1\rangle).
\end{equation}
No single-qubit states exist such that $|\Phi\rangle$ can be written as a product state:
\begin{align}
	|\Phi\rangle &\neq (\alpha_1|0\rangle + \beta_1|1\rangle)\otimes (\alpha_2|0\rangle + \beta_2|1\rangle) \\
	&= \alpha_1 \alpha_2 |0\rangle \otimes |0\rangle  +  \alpha_1 \beta_2 |0\rangle \otimes |1\rangle +  \beta_1 \alpha_2 |1\rangle \otimes |0\rangle  +  \beta_1 \beta_2 |1\rangle \otimes |1\rangle .
\end{align}
More specifically, it is not possible to find complex numbers ${\alpha_1, \alpha_2, \beta_1, \beta_2}$ such that $\alpha_1 \alpha_2 = \beta_1 \beta_2 = 1/\sqrt{2}$ and $\alpha_1 \beta_2=\alpha_2 \beta_1 = 0$. This is also known as the phenomenon of entanglement. The outcomes of measurements on both qubits of the state are perfectly correlated. Indeed, further investigation reveals that the amount of correlation exceeds the maximal classically possible amount of correlation.

\section*{The quantum non-linear solver}

In this appendix we shall outline the key working mechanisms of the quantum non-linear solver. For a detailed introduction the reader should consult \cite{QuantSolver2020}. Let us start by considering the integration of a linear differential equation system $\dot{\mathbf{x}}= \hat{A} \mathbf{x} $ first, where $\hat{A}$ is a constant operator acting on the $d$-dimensional vector $\mathbf{x}$ of dynamical variables $x_i$. By discretising the time domain and employing the forward Euler approach, we yield a set of linear equations:
\begin{equation}
\mathbf{x}\left(t_0 +  (k+1)\Delta t\right) = \mathbf{x}(t_0 +  k\Delta t) + \Delta t \hat{A} \mathbf{x}(t_0 + k \Delta t) \quad k=1\ldots T.
\end{equation}
For initial condition $\mathbf{x}(t_0) = \mathbf{x}_{ini}$, these equations can be cast in matrix form:

\begin{equation}
\begin{pmatrix}
\mathbf{1} & 0 & \ldots & &\\
-(\mathbf{1} + \Delta t \hat{A}) & \mathbf {1}&0 &\ldots& \\
0 & -(\mathbf{1} + \Delta t \hat{A}) & \mathbf {1}& 0 & \ldots \\
\vdots & 0 &\ddots& \ddots& 
\end{pmatrix} \cdot
\begin{pmatrix}
\mathbf{x}_0 \\
\mathbf{x}_1 \\
\mathbf{x}_2 \\
\vdots
\end{pmatrix} =
\begin{pmatrix}
\mathbf{x}_{ini} \\
0 \\
0 \\
\vdots
\end{pmatrix}.
\end{equation}
The history state vector $(\mathbf{x}_0,\mathbf{x}_1,\ldots,\mathbf{x}_T)$ encodes the discretised evolution of the dynamical variables. It can be derived by inverting the matrix on the left hand side. Quantum computers can accomplish this task using the Harrow-Hassedim-Lloyd (HHL) quantum matrix inversion algorithm (cf.~\citet{Harrow2009}). This quantum algorithm provides an exponential speed-up over classical algorithms and therefore allows for a resource-efficient integration of linear differential equations.

In order to solve non-linear differential equations, we generalise the above scheme to non-linear forward Euler discrete steps. Consider a differential equation system $\dot{\mathbf{x}}= \hat{f}(\mathbf{x}) \mathbf{x} $ where the operator $\hat{f}$ is now a function of the dynamical variables $\mathbf{x}$. In order to implement $T$ non-linear forward Euler equations 
\begin{equation}
\mathbf{x}\left(t_0 +  (k+1)\Delta t\right) = \mathbf{x}(t_0 +  k\Delta t) + \Delta t \hat{f}(\mathbf{x}(t_0 + k \Delta t)) \mathbf{x}(t_0 + k \Delta t),
\end{equation}
we use mean-field non-linearities. Take $N$ identical copies of a (quantum) state $X_k = \mathbf{x}(t_0 +  k\Delta t) \otimes \ldots\otimes\mathbf{x}(t_0 +  k\Delta t) \equiv \mathbf{x}(t_0 +  k\Delta t) ^{\otimes N}$. One can construct an operator $\hat{F}$ acting on these $N$ copies that upon projection on a single copy will reduce to the desired operator $\hat{f}$ within order $\mathcal{O}(\Delta t^2)$. Consequently, the matrix inversion that will be carried out by the quantum computer reads:
\begin{equation}
\begin{pmatrix}
\mathbf{1} & 0 & \ldots & &\\
-(\mathbf{1} + \Delta t \hat{F}) & \mathbf {1}&0 &\ldots& \\
0 & -(\mathbf{1} + \Delta t \hat{F}) & \mathbf {1}& 0 & \ldots \\
\vdots & 0 &\ddots& \ddots& 
\end{pmatrix} \cdot
\begin{pmatrix}
\mathbf{x}_0^{\otimes N} \\
\mathbf{x}_1^{\otimes N} \\
\mathbf{x}_2^{\otimes N} \\
\vdots
\end{pmatrix} =
\begin{pmatrix}
\mathbf{x}_{ini}^{\otimes N} \\
0 \\
0 \\
\vdots
\end{pmatrix}.
\end{equation}
After inverting the matrix on a quantum computer, we yield a history state of $N$ copies of the system state. The dynamical variables can then be evaluated by measuring individual copies.

%%%%%%%%%%%%%%%%%%%%%%%%%%%%%%%%%%%%%%%%%%%%%%%%%%%%%%%%%%%%%%%%%%%%%
% REFERENCES
%%%%%%%%%%%%%%%%%%%%%%%%%%%%%%%%%%%%%%%%%%%%%%%%%%%%%%%%%%%%%%%%%%%%%
%\bibliography{mybibliography}

%\begin{thebibliography}{long-name}
%	\bibitem{Nielsen2010} M. A. Nielsen and I. L. Chuang; Quantum Computation and Quantum Information: 10th Anniversary Edition, 10th ed. (CUP, 2011).
%	\bibitem{Nimbe2021} P. Nimbe, B.A. Weyori and A.F. Adekoya; Models in quantum computing: a systematic review. Quant. Inf. Proc. 20, 80 (2021).
%	\bibitem{Bharti2022} K. Bharti, et al.; Noisy intermediate-scale quantum algorithms. Rev. Mod. Phys, 94,015004 (2022).
%	\bibitem{Montanaro2016} A. Montanaro; Quantum algorithms: an overview. npj Quantum Inf. 2, 15023 (2016). 
%	\bibitem{Cerezo2021} M. Cerezo, et al. Variational quantum algorithms. Nat. Rev. Phys. 3, 625–644 (2021).
%	\bibitem{QuantSolver2020} Lloyds et al.
%	\bibitem{Leyton2008} S. Leyton and T. J. Osborne; A quantum algorithm to solve nonlinear differential equations.  	arXiv:0812.4423 (2008).
%	\bibitem{QuantSolver2022} Tennie, to be published
%\end{thebibliography}

\bibliographystyle{ametsocV6}
\bibliography{references}

\end{document}